\documentclass[journal]{IEEEtran}
\usepackage{graphicx}
\usepackage{subfig}

\begin{document}
\title{Proton recoil telescope based on diamond detectors for measurement of fusion neutrons}
%
%

\author{ Barbara Caiffi, Mikhail Osipenko, Marco Ripani, Mario Pillon and Mauro Taiuti   

\thanks{Manuscript received April 6, 2015. This work was supported by the Istituto Nazionale di Fisica Nucleare INFN-E strategic program.}
\thanks{Barbara Caiffi and Mauro Taiuti are with Universit\`a di Genova, Genoa, 16146 Italy. (e-mail: caiffi@ge.infn.it).}
\thanks{Mikhail Osipenko and Marco Ripani are with the Istituto Nazionale di
Fisica Nucleare, Genoa, 16146 Italy.}
\thanks{Mario Pillon is with the ENEA, Frascati, 00044 Italy.}%
}

\maketitle
\pagestyle{empty}
\thispagestyle{empty}

\begin{abstract}
Diamonds are very promising candidates for the neutron diagnostics in harsh environments such as fusion reactor. In the first place this is because of their radiation hardness, exceeding that of Silicon by an order of magnitude. Also, in comparison to the standard on-line neutron diagnostics (fission chambers, silicon based detectors, scintillators), diamonds are less sensitive to $\gamma$ rays, which represent a huge background in fusion devices. Finally, their low leakage current at high temperature suppresses the detector intrinsic noise.
In this talk a CVD diamond based detector has been proposed for the measurement of the 14 MeV neutrons from D-T fusion reaction.
The detector was arranged in a proton recoil telescope configuration, featuring a plastic converter in front of the sensitive volume in order to induce the (n,p) reaction. The segmentation of the sensitive volume, achieved by using two crystals, allowed to perform measurements in coincidence, which suppressed the neutron elastic scattering background. A preliminary prototype was assembled and tested at FNG (Frascati Neutron Generator, ENEA), showing promising results regarding efficiency and energy resolution. 
\end{abstract}

\begin{IEEEkeywords}
diamond detectors, neutronics, fusion diagnostics.
\end{IEEEkeywords}

\section{Introduction}
%
%
%
%
\IEEEPARstart{N}{eutron} diagnostics plays a particularly important role in fusion devices. In particular, in the new generation reactors such as ITER, in which huge neutron fluxes produced by the at fusion power of 500 MW must be monitored to estimate the reactor structural damage and the activation.
These diagnostics will have to withstand harsh environments, high temperatures and high EM fields. A promising technology for the neutron spectrometer is the CVD diamond-based detector.
Diamonds are semiconductor detectors, in which the charge produced by the passage of an ionizing particle in the sensitive volume is collected by means of an applied electric bias field. Diamond features a large electron/hole mobility, 2000 $\mu m^2/ (Vs)$, so the charge collection is very fast (a few ns). This makes them suitable for high rate measurements, mandatory in the high flux environment. 

Since diamond band-gap is as wide as 5.5 eV, the dark current noise at room temperature is negligible. Therefore, the detector does not require any ad hoc cooling system and can operate in a high temperature environment without intrinsic noise. Another great advantage is its compactness: a fully equipped detector may have linear size $<1$ cm. Diamond has an octahedral crystalline structure, very strongly bound. It is a low $Z$ crystal with the highest melting point, $3800^o K$, and with the largest displacement energy, 43 eV, among the ones used for neutron detection. This makes it possible to create compact devices suitable for very high radiation environments.

Diamond detectors for radiation monitoring are not a new technology: the first measurements with natural diamonds go back to 1950s \cite{CVD_hof}. Natural diamonds had been used in the past in various fields. In the 90's, they were also installed for 14-MeV neutron diagnostic in the main fusion devices, like JET \cite{NDD_Pillon} \cite{NDD_kra_JET} \cite{NDD_kra_JET2}  and TFTR \cite{NDD_kra_TFTR}.
The use of natural diamonds, however, presented some difficulties related to the limited availability  of mono crystalline diamond stones of high electronic quality, which of course, made their price inaccessible. In particular, only IIa type natural diamonds featured an electronic grade suitable for particle detection.

From 1988, artificial diamond films produced by chemical vapor deposition (CVD) were developed, making it possible to  fabricate in a reproducible manner crystals exhibiting purity and electrical properties better than those of natural IIa diamond.
CVD diamond detectors soon replaced the natural diamond detectors in the neutron diagnostics of fusion devices, getting comparable results \cite{CVD} \cite{CVD4}.

Diamond detectors can also be coated with neutron to charge particle converter, whose material has to be chosen according to the neutron energy range. For example, lithium and boron can be used in the neutron thermal range, in which the nuclear reactions  $^{6}$Li(n,$\alpha$)$t$ or $^{10}$B(n,$\alpha$)$^{7}$Li are likely, while the proton elastic recoil on plastic material can be used for high energy neutrons,  in the  1-10 MeV range. LiF coating on CVD diamonds has been used in JET tokamak \cite{LiF}. In that configuration, not only the 14-MeV neutrons had a good energy resolution through the $^{12}$C(n,$\alpha$)$^{9}$Be reaction, but also the thermal neutrons could be measured. Also a plastic converter placed in front of a natural diamond detector was used to measure the  burn up fraction in the proton recoil telescope configuration \cite{NDD_proton_recoil}. This configuration, though, gave not satisfactory results: in fact, the background due to the elastic scattering of the neutrons on the Carbon atoms compromised the measurements of the 2.5 MeV neutrons. 

In this work, a new concept of diamond neutron detector was studied. It used plastic neutron to proton converter, as in the previously cited proton recoil telescope configuration, but its sensitive volume was segmented in two diamond crystals. In this way, using fast coincidence measurements, the elastic scattering background can be rejected and, in principle, both the 14 MeV and the 2.5 MeV neutron can be measured. The Monte Carlo simulations of the spectrometer, the assembly and the measurements of its prototype will be described in details in the following sections. 

\section{Spectrometer Description and Feasibility Study}
\begin{figure}[htpb]
\begin{center}
\includegraphics[trim= 25mm 15mm 40mm 20mm ,clip=true,scale=0.5]{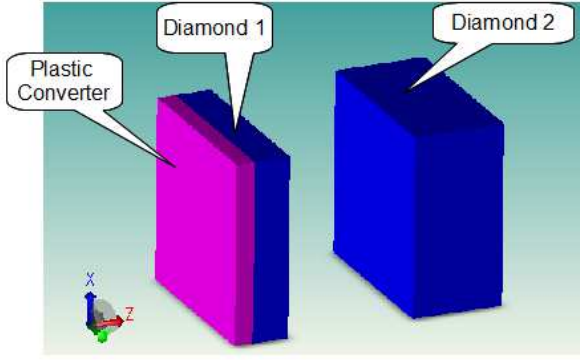}
\end{center}
\caption{Schematic representation of the proton recoil prototype.}
\label{fig:proto}
\end{figure}

The geometry of the neutron recoil telescope for 14 MeV DT neutrons studied in this thesis is shown in Fig.~\ref{fig:proto}.

The telescope was made by a plastic neutron to proton converter, located in front of the sensitive volume of the detector, forming the classical dE vs. E$_{total}$ calorimeter.
The sensitive volume of the detector was composed by two CVD diamond crystals, aligned with respect to the incoming neutron flux direction.
The recoiled protons cross one diamond and are absorbed in the second, making possible to measure fast coincidences between the signals in the two crystals. Coincidence measurements allow the rejection of the background, mainly due to elastic scattering of neutrons on carbon nuclei. This background can be particularly high because of the difference in the thickness of the converter and the diamonds and can compromise the resolution of the measurements as well as the acquisition rate in a high flux environment such as ITER tokamak.
Using an unfolding procedure, the energy of the neutrons can be reconstructed from the proton deposited energy. The use of a plastic converter makes the unfolding easier, thanks to the smoothness of the elastic scattering cross section of incoming neutron on hydrogen. The unfolding can be done also from the energy deposited by the neutron scattering on the carbon itself \cite{NDD_kra_TFTR}, but this task is complicated by various thresholds of inelastic channels and presence of resonances.
The thicknesses of the two diamonds have to  be optimized to allow 14 MeV protons to deposit enough energy in the first crystal and to stop completely in the second. The thickness of the first crystal determines the energy threshold of the protons which can be detected in coincidence. The converter thickness has to be a compromise between  a good conversion efficiency and a minimal energy loss of the recoil protons in its volume.
Furthermore, distantiating the two crystals by a known distance allows to detect in coincidence only those protons which recoil forward with respect to the incoming neutron flight direction, and thus which carry the entire neutron energy. This separation also suppresses the background of two successive elastic scatterings by a factor $\frac{L^2}{d^2}$, where $L$ is the crystal side and $d$ is the distance between the two crystals. This reduces the unfolding of the neutron spectrum to a straightforward procedure.

The neutron to proton conversion is not the only signal in coincidence though.
Other most important contributions to coincidences come from $^{12}$C(n,n'), i.e. from the elastic scattering of neutrons on the carbon nuclei in both crystals, and from the $^{12}$C(n,x$\alpha$), which are essentially $^{12}$C(n,$\alpha$)$^{9}$Be or $^{12}$C(n,3$\alpha$)n.
In the last two cases, one of the $\alpha$s can travel from one crystal to the other, to give the coincidence.



Three different prototypes were studied, all of them presented the common structure of Fig.~\ref{fig:proto}, with 3 $\times$3 mm$^2$ crystals, a 20 $\mu$m thick polyethylene converter and 50 nm thick Cr contacts. The differences between the three prototypes were in the crystal thicknesses:
\begin{itemize}
\item prot.1: thickness$_{CVD1}$=30 $\mu$m, thickness$_{CVD2}$=700 $\mu$m,
\item prot.2: thickness$_{CVD1}$=300$\mu$m, thickness$_{CVD2}$=500 $\mu$m,
\item prot.3: thickness$_{CVD1}$=100$\mu$m, thickness$_{CVD2}$=600 $\mu$m.
\end{itemize}

\begin{figure}[htpb]
\begin{center}
\includegraphics[trim= 0mm 0mm 0mm 0mm ,clip=true,scale=0.35]{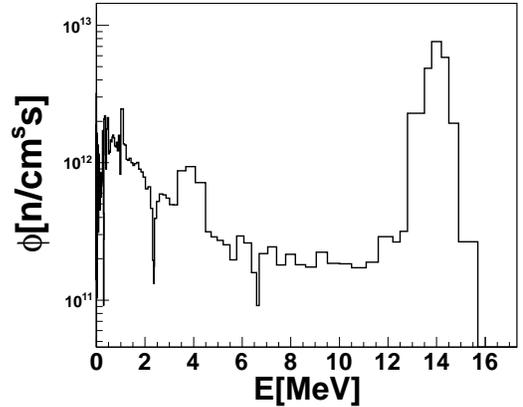}
\end{center}
\caption{Neutron spectrum expected in front of the neutron spectrometer in ITER. The 14 MeV peak has $\sigma$ about 500 keV large (T$_i$ $\sim$ 25 keV).}
\label{fig:spectrum_FW}
\end{figure}

Simulations of the detector response were performed for all the three prototypes with the Monte Carlo code Geant4 \cite{Geant4}. To this end, a realistic neutron spectrum expected after a collimator beam line which is envisaged in front of the neutron spectrometer in ITER was used (see Fig.~\ref{fig:spectrum_FW}). The 14 MeV peak of the D-T neutrons has $\sigma$ about 500 keV large, because of the thermal broadening at temperature T$_i$ $\sim$ 25 keV, which is the expected ion temperature in ITER.

\subsection{Prototype 1}

\begin{figure}[htpb]
\begin{center}
\vspace{-5pt}
  \subfloat[]{\label{fig:dEdx_idea} \includegraphics[trim= 10mm 0mm 0mm 0mm ,clip=true,scale=0.45]{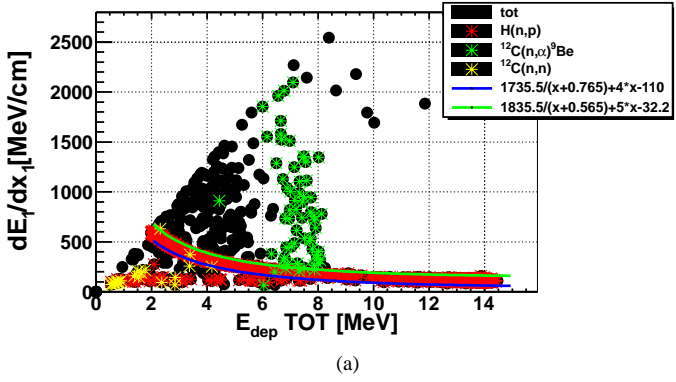}} \\       
  \subfloat[]{\label{fig:Edep_idea}\includegraphics[trim= 10mm 0mm -10mm 0mm ,clip=true,scale= 0.4]{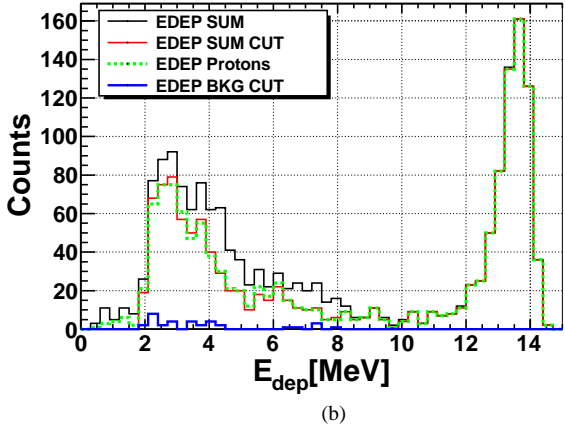}}
 \caption{(a) Correlation of energy loss in first crystal dE$_1$/dx$_1$ vs. the total energy deposited in the detector E$_{dep}$ for the prototype 1, obtained using the spectrum shown in Fig.\ref{fig:spectrum_FW}. (b) The total energy deposited in the detector by all the events (black histogram) and by all the proton events (dashed green histogram). The energy loss cut indicated by the solid lines in (a) was applied and allowed to select only the completely absorbed proton events (red histogram). The background remaining after the cut is indicated by the blue histogram.}
\vspace{-5pt}
  \label{fig:proto_idea}
  \end{center}
\end{figure}

\begin{figure}[htpb]
\begin{center}
\includegraphics[trim= 0mm 0mm 0mm 0mm ,clip=true,scale=0.4]{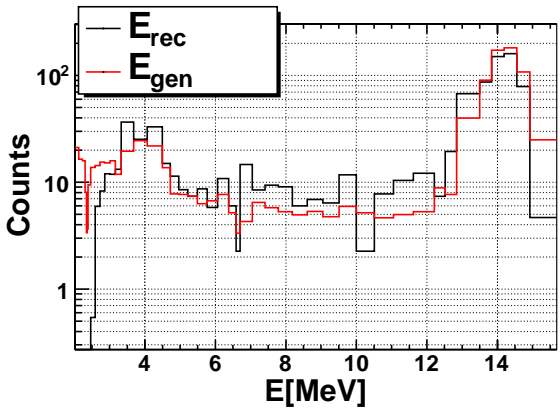}
\end{center}
\caption{Generated and reconstructed neutron energy spectra. The reconstructed energy was obtained from the proton deposited energy divided by the efficiency and adding the 300 keV, which protons lost in the non-sensitive volume of the detector.}
\label{fig:unfolding}
\end{figure}

In the  prototype 1, the first crystal was considered as thin as possible: this represented the ideal case, with the lowest neutron detection threshold. The lower limit on the crystal thickness was given by the the electronic readout noise level, which had to be lower than the signal from the proton crossing the crystal. The thickness of 30 $\mu$m was chosen assuming 30 keV equivalent noise level. The second crystal was chosen 700 $\mu$m thick, so that the sum of the thicknesses was comparable to the range of 14 MeV protons. The converter thickness was set to 20 $\mu$m, which corresponded to an energy loss of 140 keV for 14 MeV protons.

The simulations of prototype 1 response to the realistic neutron spectrum from Fig.~\ref{fig:spectrum_FW} revealed its main characteristics. 
Prototype 1 allowed to reconstruct the neutron spectrum down to 2 MeV. To discriminate the protons from the background, the energy loss cut was applied on the dE vs. E$_{total}$ correlation.
This cut is shown by the two solid lines in Fig.~\ref{fig:dEdx_idea}, where dE$_1$/dx$_1$ is obtained as the energy deposited in the first crystal divided by the crystal thickness and E$_{tot}$ is the sum of the energy deposited in the first and in the second crystal.
The total energy deposited in the crystals is shown in Fig.~\ref{fig:Edep_idea} before and after the cut. The efficiency for neutrons above 2 MeV was found to be about 4$\times$10$^{-6}$. The remaining background contamination in the reconstructed events, represented by all the events not corresponding to protons stopped completely in the telescope and remaining after the energy loss cut, was about 1\%.
The neutron spectrum was reconstructed from the deposited energy distribution, dividing it by the efficiency and adding the 300 keV, which protons lost in the non-sensitive volume of the detector. The original and reconstructed neutron spectra, shown in Fig.~\ref{fig:unfolding}, demonstrate fairly good agreement.
Prototype 1 is an extreme configuration because commercial crystals of that size (30 $\mu$m of thickness) are not available. However, it gave important indications on the intrinsic limits of this concept, like the lowest neutron reconstructed energy and the intrinsic energy resolution.

\subsection{Prototype 2}
\begin{figure}[htpb]
\begin{center}
\vspace{-5pt}
  \subfloat[]{\label{fig:dEdx_real} \includegraphics[trim= 10mm 0mm 0mm -10mm ,clip=true,scale=0.45]{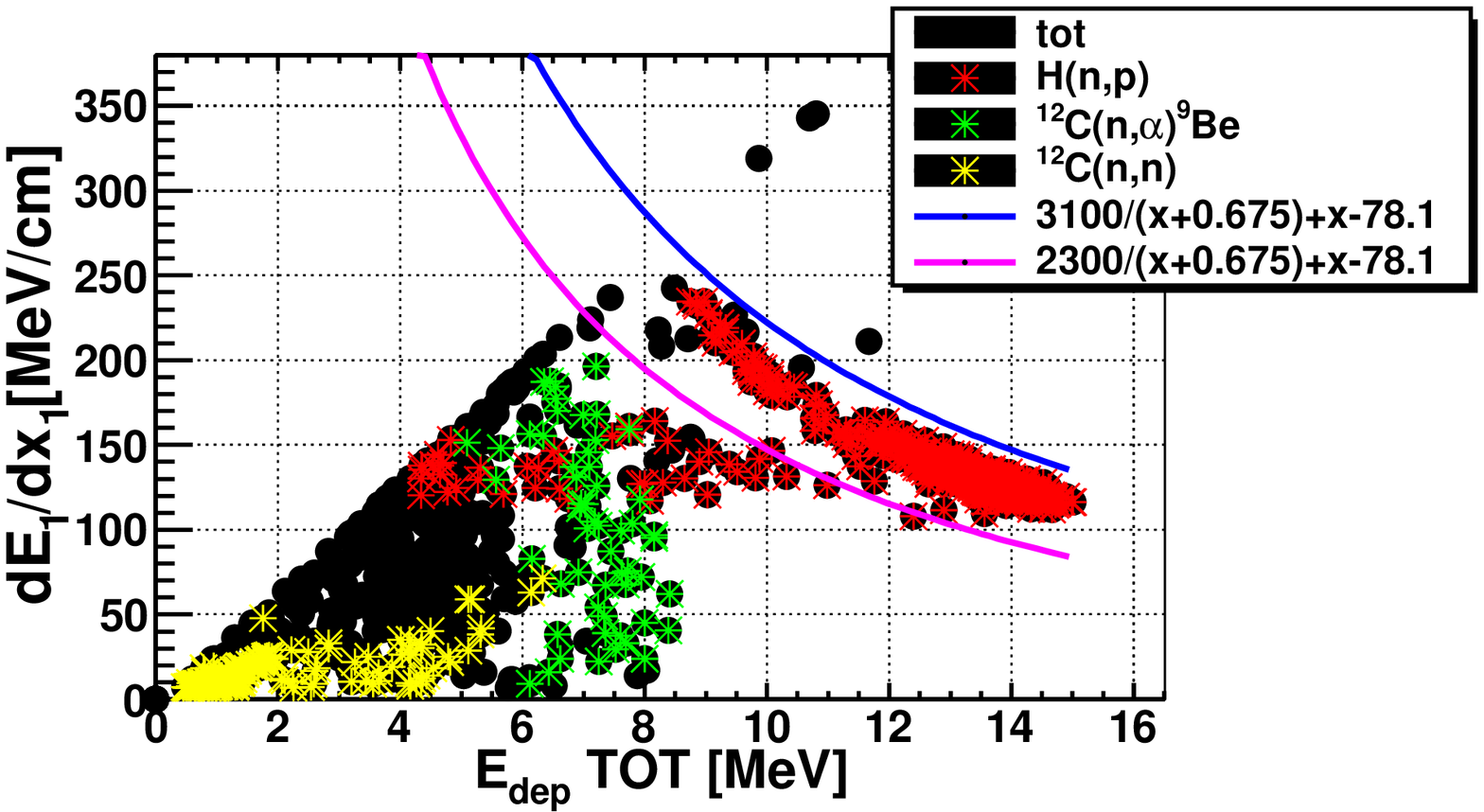}} \\        
  \subfloat[]{\label{fig:Edep_real}\includegraphics[trim= 10mm 0mm -10mm -5mm ,clip=true,scale= 0.4]{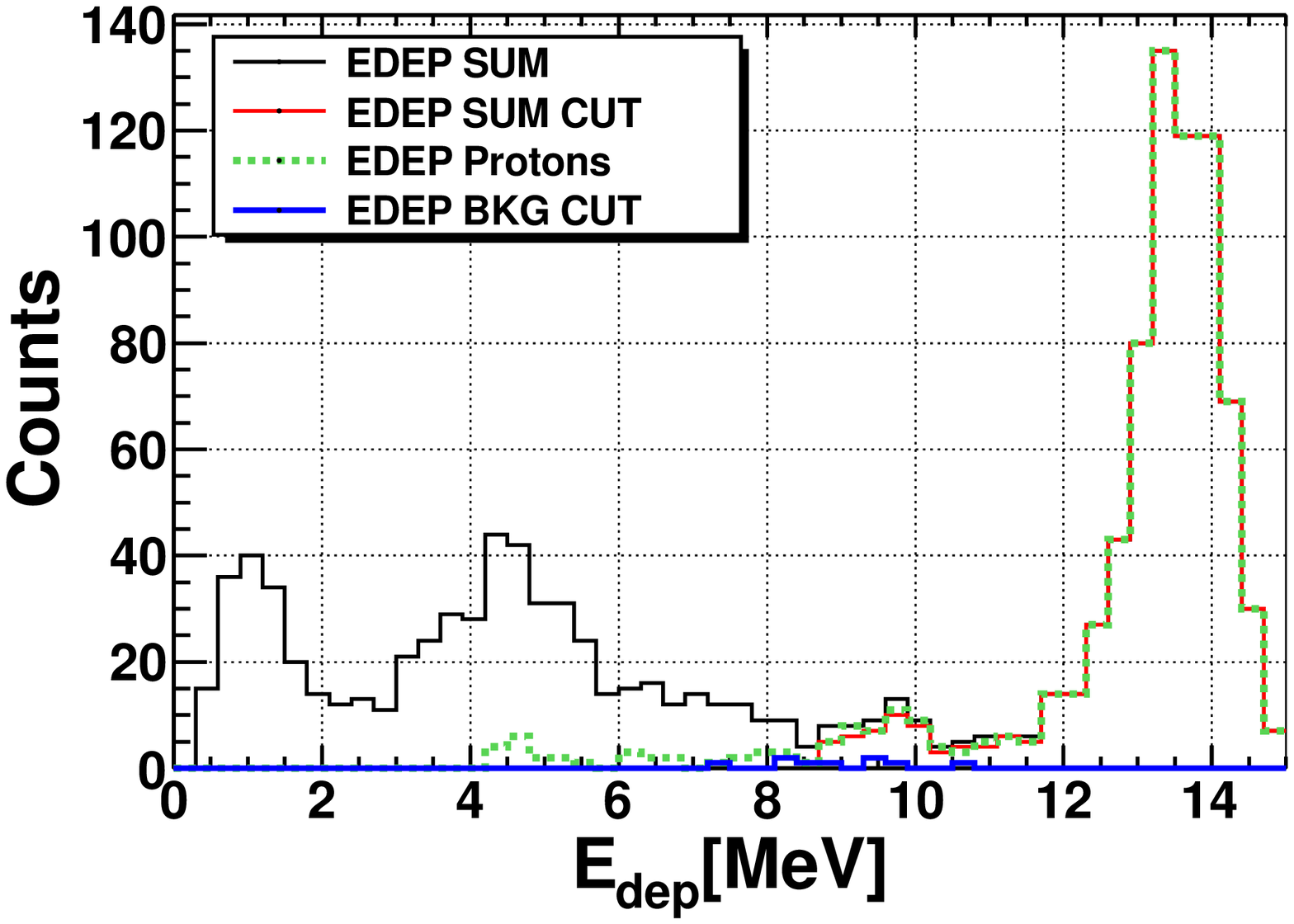}}
 \caption{The same as in Fig.~\ref{fig:proto_idea}, but for prototype 2.}
\vspace{-5pt}
  \label{fig:proto_real}
  \end{center}
\end{figure}

In the prototype 2, commercially available crystal thicknesses were chosen: the prototype was constructed with a first 300 $\mu$m thick crystal followed by a second 500 $\mu$m thick crystal. The prototype 2 allowed to reconstruct the energy spectrum of the incoming neutron down to 8 MeV.
Also in this case, an energy loss cut indicated by the solid lines in Fig.~\ref{fig:dEdx_real} was applied to select completely absorbed protons. The efficiency of this cut can be seen in the total energy deposited in the crystals, shown in Fig.~\ref{fig:Edep_real} before and after the cut. The efficiency for neutrons above 8 MeV was about 3$\times$10$^{-6}$ with the background contamination about 1\%.

\subsection{Prototype 3}
\begin{figure}[htpb]
\begin{center}
\vspace{-5pt}
  \subfloat[]{\label{fig:dEdx_int} \includegraphics[trim= 10mm 0mm 25mm 0mm ,clip=true,scale=0.5]{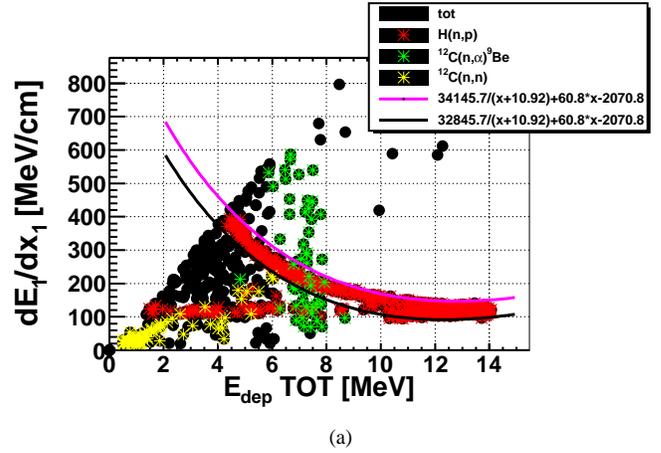}} \\        
  \subfloat[]{\label{fig:Edep_int}\includegraphics[trim= 10mm 0mm -10mm 0mm ,clip=true,scale= 0.4]{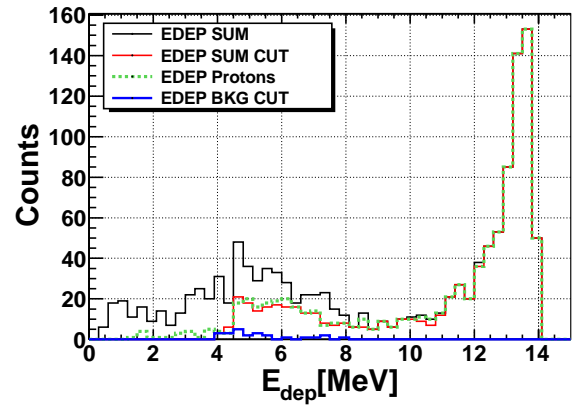}}
 \caption{The same as in Fig.~\ref{fig:proto_idea}, but for prototype 3.}
\vspace{-5pt}
  \label{fig:proto_int}
  \end{center}
\end{figure}

In the Prototype 3, crystals with thicknesses of 100 $\mu$m and 600 $\mu$m where considered, assuming that the thickness of the commercially available crystals can be reduced using the lapping or etching technique. In this case, the neutron spectrum could be reconstructed down to 4 MeV. The energy loss cut to reject all the events but the completely absorbed protons is shown in Fig.~\ref{fig:dEdx_int}. And the effect of this cut can be seen in Fig.~\ref{fig:Edep_int}. The efficiency for neutrons above 4 MeV was about 3.2$\times$10$^{-6}$, while the remaining background in the reconstructed events was less than 3\%.

In the Table~\ref{table:eff_proto}, efficiencies and remaining background contaminations are listed for the 3 prototypes considered. 

\begin{table}[htpb]
\begin{center}
\begin{tabular}{ | c | c | c | }  \hline                       
  Configuration & Efficiency & Background  \\
     & & contamination \\\hline 
  Prot.1  & 4$\times$10$^{-6}$ (E$>$2 MeV)   & 1\% \\\hline
  Prot.2  & 3$\times$10$^{-6}$ (E$>$4 MeV)   & 1\% \\\hline 
  Prot.3  & 3.2$\times$10$^{-6}$ (E$>$8 MeV) & 3\% \\\hline 
\end{tabular} 
\end{center}
\caption{\label{table:eff_proto} Efficiency and background contamination for the three different detector prototypes.}
\end{table}

In conclusion, a neutron spectrometer optimized for 14 MeV neutrons was studied. Three different prototypes with different diamond thicknesses were considered, showing that the thickness of the first diamond 
affects the threshold of the energy spectrum which can be reconstructed. In the ideal configuration (prototype 1), using a 30 $\mu$m thick crystal, the energy spectrum can be reconstructed down to 2 MeV. Using commercially available crystals, 300 $\mu$m and 500 $\mu$m, instead, the spectrum can be reconstructed only down to 8 MeV. The energy resolution depends on the thickness of the non-sensitive volumes of the detectors (contacts and converter) and in our geometry was about 270 keV. The neutron detection efficiency in all three prototypes was of the order of 10$^{-6}$ and the background contamination was of a few \%.

\section{Measurement at FNG}
\begin{figure}[htpb]
\begin{center}
\includegraphics[trim= 0mm 0mm 0mm 0mm ,clip=true,scale=0.45]{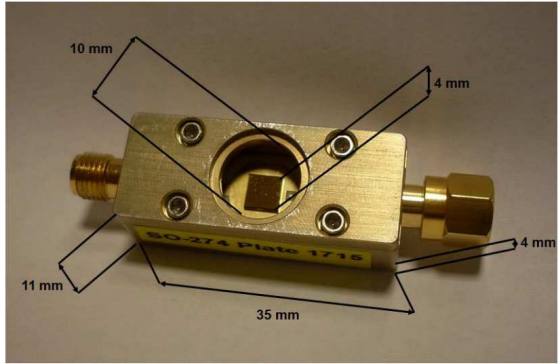}
\end{center}
\caption{Picture of DDL CVD diamond detector.}
\label{fig:DDL}
\end{figure}

\begin{figure}[htpb]
\begin{center}
 \includegraphics[trim= 0mm 0mm 0mm 0mm ,clip=true,width=0.4\textwidth]{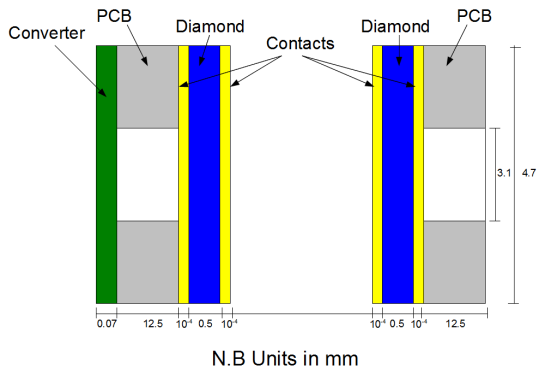}
 \caption{Schematic drawing of the prototype.}
\vspace{-5pt}
  \label{fig:proto_FNG}
  \end{center}
\end{figure}

In the previous section, the optimal geometrical parameters for the proton recoil telescope were studied. Unfortunately, the commercially available CVD diamond crystals have limited range of thicknesses. For the first exploratory measurement the standard DDL detectors~\cite{DDL} were used to test its operation principle and efficiency.
The tested telescope prototype was made by two CVD diamond detectors, arranged one in front of the other, aligned at a distance of 1.2 cm. A 70 $\mu$m thick plastic converter was placed in front of the first crystal.
One of the DDL detectors is shown in Fig.~\ref{fig:DDL}. It is composed by an electronic grade, single crystal CVD diamond, with dimensions 0.5$\times$4.7$\times$4.7 mm$^3$. It is located in a brass case, glued from one side to the PCB with a conductive glue, and bonded with micro-wires on the opposite side. Both the metallic case and the PCB are perforated, so that charged particles can travel from one crystal to the other passing only though the air.
The schematic drawing of the prototype is shown in Fig.~\ref{fig:proto_FNG}.

This prototype was tested at the Frascati Neutron Generator facility \cite{FNG}. In this facility, a deuteron beam impinges on a tritium target and up to 10$^{11}$ 14 MeV neutrons can be produced by T(d,n)$\alpha$ fusion reactions.
 
\subsection{Experimental Set-up}
\begin{figure}[htpb]
\begin{center}
\includegraphics[trim= 0mm 0mm 0mm 0mm ,clip=true,scale=0.4]{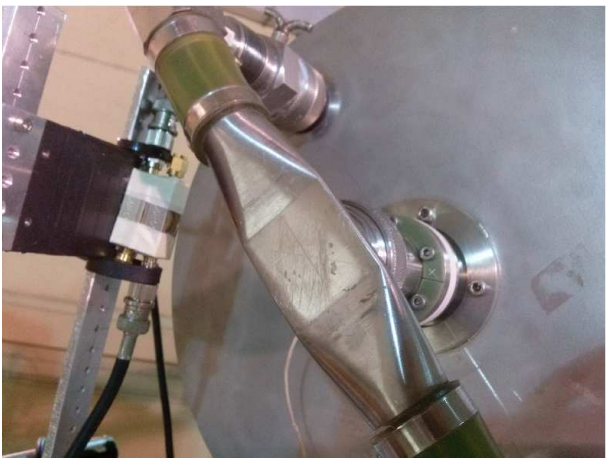}
\end{center}
\caption{Picture of the prototype installed at the FNG neutron source.}
\label{fig:foto}
\end{figure}

The prototype was fasten to a metallic support structure, at a distance of about 10 cm from the tritium target, at 90$^\circ$ with respect to the deuterium beam line. The position of the prototype with respect to the neutron source is shown in Fig.~\ref{fig:foto}.

\begin{figure}[htpb]
\begin{center}
\includegraphics[trim= 0mm 30mm 0mm 30mm ,clip=true,scale=0.4, angle=-90]{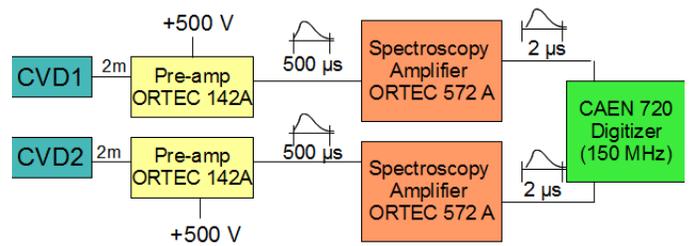}
\end{center}
\caption{Charge sensitive electronic chain.}
\label{fig:charge_chain}
\end{figure}

\begin{figure}[htpb]
\begin{center}
\includegraphics[trim= 0mm 35mm 0mm 15mm ,clip=true,scale=0.37, angle=-90]{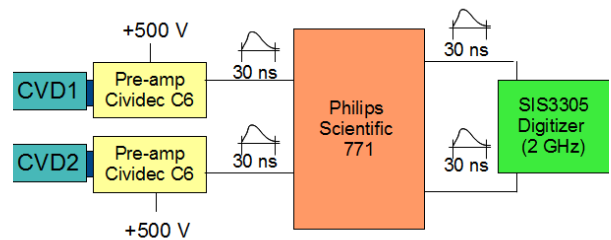}
\end{center}
\caption{Fast electronic chain.}
\label{fig:fast_chain}
\end{figure}

Two different read-out chains were used in the measurement, shown respectively in Fig.~\ref{fig:charge_chain} and Fig.~\ref{fig:fast_chain}. 
The first chain consisted of a charge sensitive pre-amplifiers, ORTEC 142A, followed by spectroscopy amplifiers, ORTEC 572A, with shaping time of 2 $\mu$s, acquired by a digitizer, CAEN DT5724, working at a sampling rate of 100 Ms/s. The two detectors were biased with +500 V and -500 V, respectively, and connected to the pre-amplifiers by 2 m long cables. The coincidence analysis was performed off-line.
The second chain consisted of trans-impedance pre-amplifiers Cividec C6, directly connected to the detectors, followed by an amplifier, Philips Scientific 771, and by a digitizer, SIS3305, working at a sampling rate of 5 Gs/s. Also in this case, voltages of +500 V and -500 V were supplied to the detectors. The firmware of the digitizer was modified, in order to acquire the data both in OR mode (trigger on either of the two detectors) and in AND mode (in coincidence).

A simulation of the prototype response to the FNG neutron flux was carried out. A detailed MCNP~\cite{MCNP} input file, describing the geometry and the neutron source was provided by ENEA and allowed to calculate the neutron flux and spectrum in the region of interest. Unfortunately, MCNP is not able to generate the products of nuclear reactions such as $^{12}$C(n,$\alpha$)$^{9}$Be or (n,p). So the simulation was made in two steps: the neutron flux and spectrum at the converter surface were evaluated using the code MCNP, while the detector response to the neutrons was simulated with Geant4. 

\begin{figure}[htpb]
\begin{center}
\includegraphics[trim= 0mm 0mm 0mm 0mm ,clip=true,scale=0.4, angle=-90, bb=100 150 550 700]{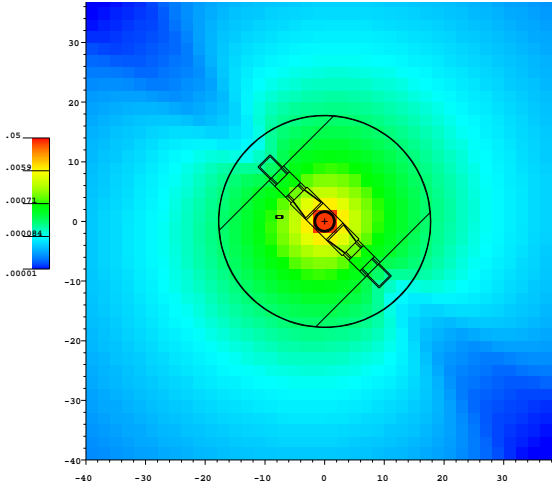}
\end{center}
\caption{2D map of the neutron flux per source particle in the transverse plane (with respect to deuteron beam). The detector boundaries, small rectangle, are located around coordinates X=-8 cm and Y=0.75 cm.}
\label{fig:FNGmap}
\end{figure}

\begin{figure}[htpb]
\begin{center}
\includegraphics[trim= 0mm 0mm 0mm 0mm ,clip=true,scale=0.5]{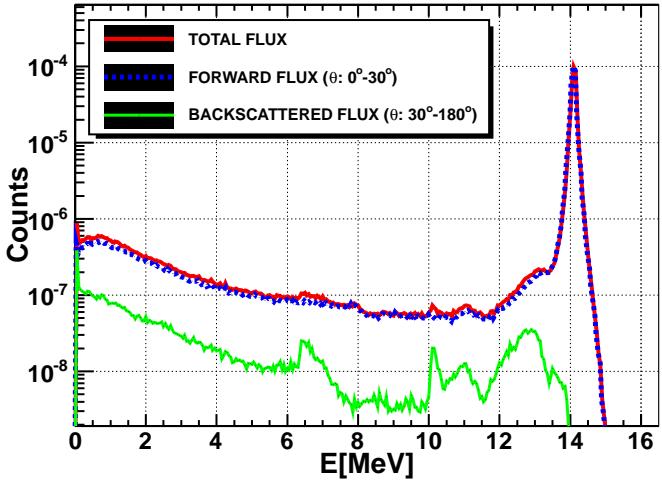}
\end{center}
\caption{Spectrum of the neutrons generated by FNG on the detector surface with its forward and large angle components. Up to 90\% of the neutrons go in the forward direction (0$^\circ$-15$^\circ$). The broadening of the 14 MeV peak is about 70 keV.}
\label{fig:FNGspectrum}
\end{figure}

In the first simulation the detector was added into the MCNP FNG input file to tally the neutron flux at its boundaries. The obtained neutron flux map per source particle is shown in Fig.~\ref{fig:FNGmap}, while the neutron spectrum at the front surface of the converter is shown in Fig.~\ref{fig:FNGspectrum} for different flight directions. Up to the 90\% of the neutrons crossed the converter surface at small angles, from 0$^o$ to 15$^o$, with respect to its normal vector, so the neutron source for the second step was approximated by a mono-directional surface source, placed just in front of the converter.
 
\subsection{Data analysis and results}
\begin{figure}[htpb]
\begin{center}
\vspace{-15pt}
\subfloat[Fast chain]{\label{fig:fast_chain_single} \includegraphics[trim= 0mm 0mm 0mm 0mm ,clip=true,width=0.4\textwidth]{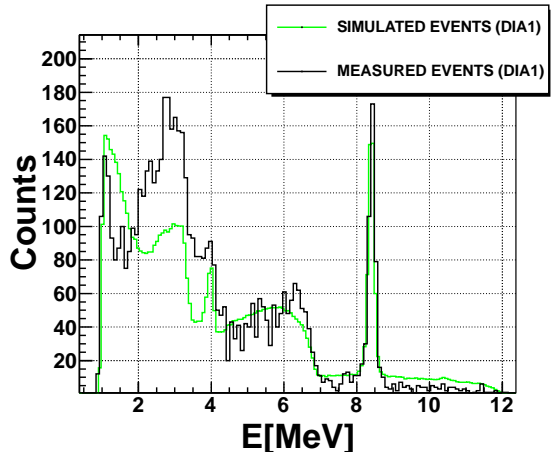}} \\        
  \subfloat[Charge sensitive chain]{\label{fig:charge_chain_single} \includegraphics[trim= 0mm 0mm 0mm 0mm ,clip=true,width=0.4\textwidth]{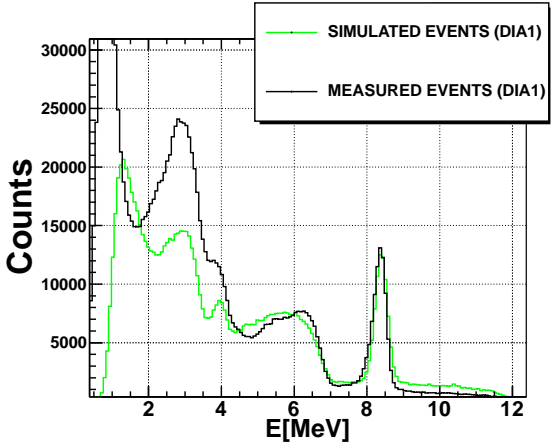}} \\         
 \caption{Energy deposited in single crystal measured with fast chain (a) and with charge sensitive chain (b). The Geant4 simulations, broadened with the electronic resolution of each chain, are also drawn. The presence of high noise during the charge sensitive chain measurements deteriorated the energy resolution. In the deposited energy spectrum, three regions can be identified: below 4 MeV there is the elastic scattering shoulder, from 4 MeV to 6 MeV there is the (n,3$\alpha$) shoulder and around 8 MeV there is the $^{12}$C(n,$\alpha$)$^{9}$Be peak. The discrepancy between measured and simulated spectra in the $<$4 MeV range could be due to the lack of some inelastic reactions with threshold $>$10 MeV in Geant4.}
\vspace{-5pt}
  \label{fig:single}
  \end{center}
\end{figure}

The energy deposited in single crystal is shown in Fig.~\ref{fig:single} in comparison with the simulations. For a better comparison, the simulated spectra were broadened with the electronics resolution ($\sigma$ $\sim$ 70 keV for the fast chain and 250 keV for the charge sensitive chain). During the acquisition with the charge sensitive chain, high electronic noise was present. This behavior could be related to the 2 m long cables used to keep the charge pre-amplifiers far from the neutron source. The high noise level compromised the energy resolution, which with this detectors could be better than 50 keV with the same charge sensitive chain.
In the deposited energy spectrum in a single detector (OR trigger), three regions can be identified:
below 4 MeV there is the elastic scattering shoulder,
then from 6 to 8 MeV there is the shoulder of the $^{12}$C(n,3$\alpha$)n reaction
and around 8 MeV there is the $^{12}$C(n,$\alpha$)$^{9}$Be peak.
Without coincidence trigger the recoil protons cannot be separated from the background, except in the range above 8 MeV $^{12}$C(n,$\alpha$)$^{9}$Be peak, where most of the events are due to proton absorption in a single crystal.
Instead, the energy deposited by recoil protons, which hit both crystals in coincidence, represented a continuous shoulder which extents up to 10 MeV in the first crystal and up to 6 MeV in the second.
The general agreement between the measured and the simulated deposited energy spectra is satisfactory. 
The discrepancy in the $<$4 MeV range could be due to the lack of some inelastic reactions with threshold $>$10 MeV in Geant4.

\begin{figure}[htpb]
\begin{center}
\includegraphics[trim= 0mm 0mm 0mm 0mm ,clip=true,scale=0.4]{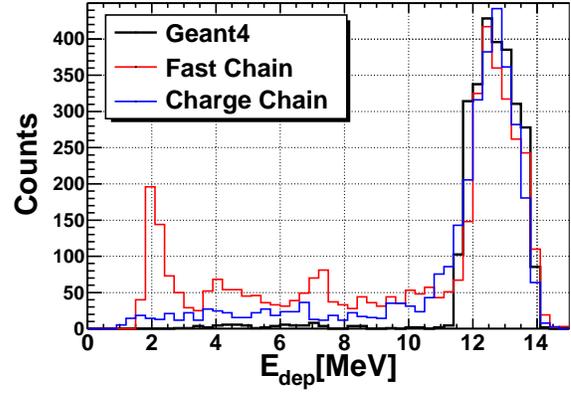}
\caption{The total energy deposited in coincidence in the two crystals measured with fast chain (red) and charge sensitive chain (blue) in comparison with Geant4 simulations (black). The width of the peak is due to the 11$^\circ$ solid angle aperture between the two crystals.}
\vspace{-5pt}
\label{fig:coinc}
\end{center}
\end{figure}

The total energy deposited in the two crystals in coincidence is shown in Fig.~\ref{fig:coinc} for both the electronic readouts and for the simulations. The results are in good agreement: in both the measurements and in the simulation the distribution of the total deposited energy exhibits a broad peak from 12 up to 14 MeV. Protons lost on average 140 keV traveling though the converter and the contacts, while electronics resolution ranged from 70 (fast chain) to 250 keV (charge sensitive chain). The peak width is mostly due to the 11$^\circ$ solid angle aperture between the two crystals, confirmed by simulations.

Tested telescope prototype efficiency to the 14 MeV neutrons in coincidence was found to be about 2.5$\times$10$^{-6}$ for both the electronic readouts and for the simulations, within 15 \%, as shown in Table~\ref{tab:FNG_eff}. In the Table, N$_{source}$ is the total number of neutrons impinging on the converter surface. For the measurements, this number was evaluated as:
$$
N_{source}=N_{\alpha} \times \epsilon_{\alpha} \times \epsilon_{MCNP} \times S_{converter}
$$
\noindent where N$_{\alpha}$ is the number of $\alpha$ particles detected by the $\alpha$ counter placed in the FNG beam line, $\epsilon_{\alpha}$=4.87$\times$10$^{7}$ is the conversion factor to obtain the total neutron yield, $\epsilon_{MCNP}$=6.3$\times$10$^{-4}$ is the neutron flux at the converter surface per FNG source particle, which was evaluated using MCNP. N$_{rec}$ is the number of events reconstructed in coincidence, applying an energy threshold of 10 MeV.
The efficiency obtained in the measurement and simulations is also compatible with the following rough estimate:
$$
\epsilon=\frac{\rho N_A}{A} n_H\Delta x \sigma_{(n,p)}(14 MeV) \frac{S_{hole}}{S_{converter}} \epsilon_{\theta}
\simeq 2.5 \times 10^{-6}
$$
\noindent where $\rho$=1 gr/cm$^3$ is the density of the polyethylene, whose chemical formula is $C_2 H_4$, A=28 is its the molar mass and n$_H$=4 is the number of H atoms in the molecule, N$_A$ is the Avogadro number, $\Delta$ x=70 $\mu$m is the converter thickness, $\sigma_{(n,p)}(E_n=14 MeV)$=0.8 b is the neutron cross section on the plastic converter, S$_{hole}$/S$_{converter}$=0.11 is the ratio between the PBC hole area and the converter surface, which represents the fraction of protons which actually can reach the crystals. $\epsilon_{\theta}$=0.015 is the solid angle fraction seen by a proton traveling from the first crystal to the second, without taking into consideration the boost in the laboratory frame due to the beam momentum. Despite the approximations of
this estimate, it gives the correct value of the prototype efficiency within 10\%. 
 
\begin{table}[htpb]
\begin{center}
\begin{tabular}{ | c | c | c | c | }  \hline                       
                   & N$_{source}$        & N$_{rec}$ (E$>$10 MeV) & Efficiency \\\hline 
Geant4 simulations & 10$^8$              & 235                    & 2.4$\times$10$^{-6}$\\\hline 
Charge Chain       & 9.1$\times$10$^{8}$ & 2489                   & 2.7$\times$10$^{-6}$\\\hline 
Fast Chain         & 10$^{9}$            & 2229                   & 2.2$\times$10$^{-6}$\\\hline
\end{tabular}
\end{center}
\caption{\label{tab:FNG_eff}Telescope prototype efficiency simulated and measured at FNG.}
\end{table}

The tested assembly with commercial detectors allowed to perform a feasibility measurement, useful to prove the validity of the concept and to estimate the efficiency and the resolution of the detector. Though, a customized prototype would be required to investigate further the feasibility of a neutron spectrometer for lower energy.
In fact, in our measurements, only protons with energy higher than 10 MeV were detected in coincidence, because the first crystal, 500 $\mu$m thick, stopped the lower energy protons. For this reason, the reconstruction of the neutron spectrum was limited to the [10-14 MeV] window.
Using crystals with different thicknesses would allow to reconstruct
the neutron spectrum down to 2 MeV, as shown in the previous section. 
Furthermore,
increasing the distance between detectors would improve the resolution without any unfolding procedure.

\section{Conclusions}
In this work, a new concept of proton recoil telescope was studied and realized. The prototype was optimized for 14 MeV neutrons from D-T fusion reaction. It was designed with a plastic converter in front of the sensitive volume, made by two CVD diamonds, placed at some distance. Using two crystals allows to perform coincidence measurements, which greatly reduce the background due to elastic scattering of the neutrons on the carbon nuclei. Distantiating the crystals allows to select only the protons recoiled forward with respect to the incoming neutron direction, and thus of the same energy. Simulations of the detector response were carried out with the Monte Carlo code Geant4, to optimize the geometrical parameters such as the converter and the crystal thickness, their relative distance, etc. Three different prototypes were considered, which accounted for different diamond thicknesses: a first ideal prototype, composed by a first very thin crystal, 30 $\mu$m thick, and a second crystal 700 $\mu$m thick, placed at 1 cm of distance with respect to the first, was studied. The second prototype used commercially available diamond thicknesses, 300 $\mu$m and 500 $\mu$m respectively, also at 1 cm of distance. The third prototype considered an intermediate situation, assuming that the lapping or etching technique could reduce the thickness of the commercial crystals. For this third configuration, 100 $\mu$m and 600 $\mu$m thick crystals at 1 cm of distance were considered. Very promising results came from those simulations, showing satisfactory efficiency, of the order of 3$\times$10$^{-6}$, and a rejection of the competitive process contamination to a few \% for all the prototypes studied. In the prototype 1, the incoming neutron spectrum could be reconstructed down to 2 MeV, so in principle it represented the best option. Though, 30$\mu$m thick crystals are not commercially available and their production seems to be difficult. For prototype 2 and 3, whose assembly presents fewer difficulties, an energy threshold of 4 MeV and 8 MeV, respectively, was imposed by the thickness of the first crystal.
A preliminary prototype was assembled using standard DDL detectors with dimensions equal to 500 $\mu$m of thickness and 4.7$\times$4.7 mm$^{2}$ of area. Despite the crystal thicknesses were not optimized, this preliminary prototype allowed to prove the operation principle of the concept.
Such prototype was tested at the Frascati Neutron Generator facility. The detector response to the neutron flux was analyzed and compared with Geant4 simulations, which exhibited good agreement with. The prototype featured an efficiency to the 14 MeV neutron detection in coincidence of about 2.5$\times$10$^{-6}$ and a resolution of about 70 keV for the fast chain and 250 keV for the charge chain, which was affected by large electronic noise. Because of the thickness of the first crystal, the neutron spectrum was reconstructed only down to 10 MeV.
Given the promising results of the measurements, a prototype with custom crystals, of optimized thicknesses found in our study, should be fabricated and tested to demonstrate the detector performances for lower energy neutrons and allow the spectrum unfolding in the $>$2 MeV energy range.


\section*{Acknowledgment}
Authors would like to acknowledge excellent support provided during the experiment by the ENEA FNG  facility staff. We also want to thank Stefano Loreti for technical and electronic support. 


%

\end{document}